\documentstyle[sprocl]{article}

\def\be{\begin{equation}}
\def\ee{\end{equation}}
\def\bea{\begin{eqnarray}}
\def\eea{\end{eqnarray}}

\begin{document}

\title{\hfill OKHEP--98--11\\
SONOLUMINESCENCE AND THE DYNAMICAL CASIMIR EFFECT}

\author{K. A. MILTON}

\address{Department of Physics and Astronomy, The University of
Oklahoma\\Norman, OK 73019-0225, USA\\E-mail: milton@mail.nhn.ou.edu}


\maketitle\abstracts{It has been suggested by various authors
that the `dynamical Casimir effect' might prove responsible
for the production of visible-light photons in the bubble
collapse which occurs in sonoluminescence.  Previously, I have
argued against this point of view based on energetic considerations,
in the adiabatic approximation.  Those arguments have recently been
strengthened by the demonstration of the equivalence between van
der Waals and Casimir energies.  In this note I concentrate on
the other extreme possibility, that of the validity of the `sudden
approximation'  where in effect the bubble instantaneously ceases to
exist.
Previous estimates which seemed to support the relevance of the
Casimir effect are shown to be unconvincing because they
require macroscopic changes on excessively small time scales,
involving
the entire volume of the bubble at maximum radius.}

\section{Introduction}
The production mechanism\footnotetext{Talk at the Fourth Workshop
on Quantum Field Theory Under the Influence of External Conditions,
Leipzig, 14--18 September, 1998} of the intense flashes of light 
which occur at the end of
bubble collapse in sonoluminescence remains mysterious.\cite{review}
A particularly intriguing possibility, put forth by Schwinger,
was that the Casimir effect in some dynamical manifestation was
responsible.\cite{js1,js2,js3,js4,js5}  This idea was extended first
by Eberlein,\cite{eberlein} and later by Carlson, Liberati, and
others.\cite{carlson,visser}

Let us start by reviewing the relevant numbers for sonoluminscent
light emission.  Typically, a bubble of air in water is held in the
node of an acoustic standing wave with overpressure of about 1
atmosphere,
of a frequency 20 kHz. The bubble goes from a maximum radius of $\sim
4\times
10^{-3}$ cm to a minimum radius of $\sim4\times 10^{-4}$ cm with a
time
scale $\tau_c$ of $10^{-5}$ s.  The flash of light, which occurs near
minimum
radius, has a time scale $\tau_f$ of less than $10^{-11}$ s, and is
characterized by the emission of $10^6$ optical photons, so about
$10$ MeV of
light energy is emitted per flash.

It seems likely that the adiabatic approximation should be valid.  If
the
flash scale is not orders of magnitude less than $10^{-11}$ s, that
scale is
long compared to the optical time scale, $\tau_o\sim10^{-15}$ s.  In
that case,
we can immediately test the Casimir idea.  The Casimir energy of a
dielectric
ball in vacuum is equivalent to that of a bubble in a dielectric
medium, and
has recently been definitively evaluated.\cite{brevik,barton}  The
Casimir
energy of a dilute ball, of dielectric constant $\epsilon$,
$|\epsilon-1|\ll1$, of radius $R$
is
\begin{equation}
E={23\over1536\pi R}(\epsilon-1)^2,
\label{casball}
\end{equation}
which may be alternatively calculated by summing the van der Waals
energies
between the molecules that make up the medium.\cite{milton}
This value is 10 orders of magnitude too small to be relevant, as
well
as of the wrong sign.  This is hardly a surprising result, since the
magnitude
of the effect is what one would expect from dimensional
considerations.

However, others have come to an opposite conclusion. In particular,
Schwinger,
\cite{js2} without relying on detailed calculations, asserted that
the
`dielectric energy, relative to that of the vacuum' was
\begin{equation}
E_c=-\int{(d{\bf r})(d{\bf k})\over(2\pi)^3}{1\over2}
k\left(1-{1\over
\epsilon({\bf r})^{1/2}}\right).
\label{casbulk}
\end{equation}
Although he argued this was true for slow variation in the dielectric
constant, he applied it to a hole of radius $a$ with a dielectric
medium,
therefore with a  discontinuous boundary:
\begin{equation}
E_c={R^3\over12\pi}K^4\left(1-{1\over\epsilon^{1/2}}\right).
\label{casbulk2}
\end{equation}
Here $K$ represents an ultraviolet cutoff, which
Schwinger took to be $K\sim 2\times 10^5$ cm$^{-1}$, which gives a
sufficient energy, $E_c\sim 6$ MeV, to be relevant.

This conclusion is supported by the work of Carlson et
al.,\cite{carlson}
who obtain the identical result.
Why is there a discrepancy of the conclusion of these authors
 with the result given in Eq.~(\ref{casball})?
The answer is simple.  The term that Schwinger \cite{js2} and Carlson
et
al.\cite{carlson} keep is indeed present as a quartically divergent
term if one simply sums normal modes.  But this is a intrinsic
contribution
to the self-energy of the dielectric medium.  It was quite properly
subtracted off at the outset in the first paper on the Casimir energy
of
a dielectric ball,\cite{me} as it was in Schwinger's own detailed
papers
on the Casimir effect.\cite{jscas} A detailed analysis of this issue
is given
in Ref.~\cite{milton}.  As Barton has noted, such divergent
volume and surface terms
`would be combined with other contributions to the bulk and to the
surface
energies of the material, and play no further role if one uses the
measured values.' \cite{barton} In other words, they serve to
renormalize
the phenomenological parameters of the model.

Further support for the irrelevance of the bulk energy comes from the
above-noted identity between the dilute Casimir energy and the van
der
Waals energy.\cite{brevik,barton,milton}
  This would seem {\it prima facie\/} evidence that the
finite remainder is unambiguously determined.  Note that the summed
van der
Waals energy must go like $(\epsilon-1)^2$, not the $\epsilon-1$
behavior
that Eq.~(\ref{casbulk}) displays.

\section{Acceleration and Temperature}

It seems plausible that the dynamical Casimir effect is closely
allied
with the so-called Unruh effect, \cite{unruh} wherein an accelerated
observer, with acceleration $a$, sees a bath of photons with
temperature $T$,
\begin{equation}
T={a\over2\pi}.
\label{unform}
\end{equation}
Indeed, the observed radiation in sonoluminescence is consistent with
the tail of a blackbody spectrum, with temperature $\sim$20,000
K.\footnote{The
temperature may be even higher.  If so, $\tau_f$ is correspondingly
reduced.}
  That is,
$kT$ is about 1 eV.  Let us, rather naively, apply this to the
collapsing
bubble, where $a=d^2 R/dt^2\sim R/\tau_f^2$, where $\tau_f$ is some
relevant
time scale for the flash.  We then have
\begin{equation}
kT\sim{R\over(c\tau_f)^2}\hbar c,
\end{equation}
or
\begin{equation}
1 \,\mbox{eV}\sim {10^{-3} \mbox{cm}\, 2\times 10^{-5}
\mbox{eV-cm}\over
\tau_f^2(3\times
10^{10}\mbox{cm\,s}^{-1})^2}\sim{10^{-29}\mbox{eV}\over\tau_f^2
(\mbox{s}^2)}.
\end{equation}
That is, $\tau_f\sim10^{-15}$ s, which seems implausibly short; it
implies a
characteristic velocity $R/\tau_f\sim10^{12}$ cm/s $\gg c$.  It is
far shorter
than the upper limit to the flash duration, $10^{-11}$ s.  Indeed, if
we
use the latter in the Unruh formula (\ref{unform}) we get a
temperature
about 1 milli Kelvin!  This conclusion seems consistent with that of
Eberlein,\cite{eberlein} who indeed stressed the connection with
the Unruh effect, but whose numbers required superluminal velocities.

However, we must remain open to the possibility that discontinuities,
as in a shock, could allow changes on such short time scales without
requiring superluminal speeds.  Indeed, Liberati et al.,\cite{visser}
following Schwinger's earlier suggestion,\cite{js1,js3} indeed assume
an extremely short time scale, so that rather than the adiabatic
approximation
discussed above being valid, a sudden approximation is more
appropriate.
We therefore turn to an analysis of that situation.

\section{Instantaneous collapse and photon production}

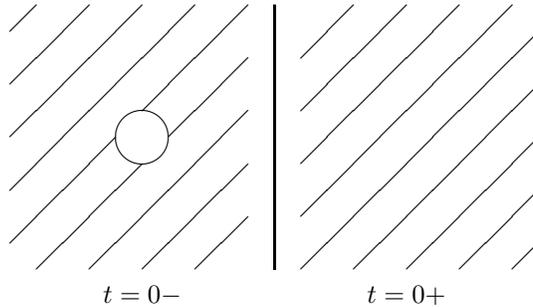
\begin{figure}
\centering
\begin{picture}(200,100)
\thicklines
\put(100,0){\line(0,1){100}}
\thinlines
\put(50,50){\circle{20}}
\put(110,0){\line(1,1){90}}
\put(130,0){\line(1,1){70}}
\put(150,0){\line(1,1){50}}
\put(170,0){\line(1,1){30}}
\put(190,0){\line(1,1){10}}
\put(110,20){\line(1,1){80}}
\put(110,40){\line(1,1){60}}
\put(110,60){\line(1,1){40}}
\put(110,80){\line(1,1){20}}
\put(70,0){\line(1,1){20}}
\put(50,0){\line(1,1){40}}
\put(30,0){\line(1,1){60}}
\put(10,0){\line(1,1){40}}
\put(60,50){\line(1,1){30}}
\put(0,10){\line(1,1){40}}
\put(50,60){\line(1,1){40}}
\put(0,30){\line(1,1){70}}
\put(0,50){\line(1,1){50}}
\put(0,70){\line(1,1){30}}
\put(0,90){\line(1,1){10}}
\put(50,-10){\makebox(0,0){$t=0-$}}
\put(150,-10){\makebox(0,0){$t=0+$}}
\end{picture}
\caption{The sudden collapse of an otherwise static bubble.}
\label{fig1}
\end{figure}

The picture offered by Liberati et al.\cite{visser} is that of the
abrupt
disappearance of the bubble at $t=0$, as shown in Fig.~\ref{fig1}.
On the face of it, this picture seems preposterous---the bubble
simply
disappears
and water is created out of nothing.  It is no surprise that a large
energy
release would occur in such a case.  Further, the static Casimir
effect
calculations employed in Ref.~\cite{visser} are invalid in this
instantaneously
changing model.  Therefore, rather than computing Bogoliubov
coefficients
from the overlap of states belonging to two static configurations,
let us
follow the original methodology of Schwinger,\cite{js1,js3}
which is essentially equivalent.

As in Schwinger's papers, let us confine our attention to the
electric (TM)
modes. They are governed by the time-dependent Green's function
satisfying
\begin{equation}
(\partial_0\epsilon(x)\partial_0-\nabla^2)G(x,x')=\delta(x-x').
\end{equation}
The photon production is given by the effective two-photon source
\begin{equation}
\delta(JJ)=i\delta G^{-1}=i\partial_0\delta\epsilon(x)\partial_0.
\label{jj}
\end{equation}
The effectiveness for producing a photon in the momentum element
centered
about $\bf k$ is
\begin{equation}
J_k=\sqrt{{(d{\bf k})\over(2\pi)^3}{1\over2\omega}}\int(dx)
e^{-i({\bf k\cdot
r}-i\omega t)}J(x),\quad \omega=|{\bf k}|.
\label{jk}
\end{equation}

Let us follow Schwinger and consider one complete cycle of
disappearance
and re-appearance of the bubble, which we assume disappears for a
time $\tau_c$:
For a bubble centered at the origin, the dielectric constant as a
function
of time within the volume of the bubble is then taken to be
\begin{equation}
r<R:\quad \epsilon(r)=1+(\epsilon'-1)\eta(\tau_c/2-|t|).
\end{equation}
Here $\epsilon'$ is the dielectric constant of all space when the
bubble is gone.  The dielectric constant of the region outside the
volume occupied by the bubble is
\begin{equation}
r>R:\quad
\epsilon(r)=\epsilon+(\epsilon'-\epsilon)\eta(\tau_c/2-|t|).
\end{equation}
Here $\epsilon$ is the dielectric constant outside the bubble when
the
bubble is present.
Occurring here is the unit step function,
\begin{equation}
\eta(x)=\left\{\begin{array}{cc}
1,&x>0,\\
0,&x<0.\end{array}\right.
\end{equation}
Clearly, this model is based on the assumption
that the disappearance time is short compared to the complete cycle
time of bubble collapse and re-expansion.

In the spirit of a first approximation, let us suppose all the
dielectric
constants are nearly unity, that is, that we are dealing with dilute
media.  Let us further assume, appropriate to the instantaneous
approximation,
that the medium is a gas, which is capable of instantaneously filling
the
bubble.  Then because the deviation of the dielectric constant from
unity is proportional to the matter number density $N$,
\begin{equation}
\epsilon-1=4\pi N\alpha,
\end{equation}
where $\alpha$ is the {\it constant\/} molecular polarizability,
matter conservation implies
\begin{equation}
(\epsilon'-1)V=(\epsilon-1)(V-v),
\end{equation}
where $V$ is the volume of all space, and $v$ is the volume of the
bubble.
Thus the change of the dielectric constant inside the bubble, and
outside,
respectively, is
\begin{eqnarray}
\delta\epsilon_{\rm in}&=&(\epsilon'-1)\eta(\tau_c/2-|t|),\nonumber\\
\delta\epsilon_{\rm out}&=&(\epsilon'-\epsilon)\eta(\tau_c/2-|t|)
=-(\epsilon'-1){v\over V-v}\eta(\tau_c/2-|t|).
\end{eqnarray}
The latter term here appears to be very small, and was therefore
disregarded
in Ref.~\cite{js1,js3,visser}.  However, we will see that the
inclusion of
this term could be significant.

{}From Eqs.~(\ref{jj}) and (\ref{jk}), the two-photon production
amplitude is
proportional to ($v\ll V$)
\begin{eqnarray}
J_kJ_{k'}&=&\sqrt{{(d{\bf k})\over(2\pi)^3}{(d{\bf k'})\over(2\pi)^3}
{1\over2\omega2\omega'}}\int(d{\bf r'})\int_{-\tau/2}^{\tau/2}
 dt \,e^{-i({\bf k+k')\cdot
r}+i(\omega+\omega')t}(-i\omega\omega')\nonumber\\
 &&\times(\epsilon'-1)\left[\eta(a-r)-{v\over
V}\eta(r-a)\right]\nonumber\\
 &\propto&(\epsilon'-1)\int_{-\tau/2}^{\tau/2}
dt\,e^{i(\omega+\omega')t}
 (-i\omega\omega')\bigg[\int_{\rm in}(d{\bf r})e^{-i({\bf k+k')\cdot
r}}
 \nonumber\\
 &&\qquad\mbox{}-{v\over V}
 \int_{\rm out}(d{\bf r})e^{-i({\bf k+k')\cdot r}}\bigg].
 \label{twophoton}
 \end{eqnarray}
 The probability of emitting two photons is proportional to the
square
 of this amplitude.  For sufficiently short wavelengths, $\lambda\ll
R$,
 the square of the quantity in square brackets in
Eq.~(\ref{twophoton})
 is the product of $(2\pi)^3\delta({\bf k+k'})$ and $v$, that is, if
the
 exterior contribution is negligible,
 \begin{equation}
 |J_kJ_{k'}|^2\propto(\epsilon'-1)^2\omega^2\sin^2\omega\tau_c\,
 \delta({\bf k+k'})v.
 \label{prob}
 \end{equation}
This is the same result found by Schwinger,\cite{js1,js3} and by
Liberati
et al.\cite{visser}  However, if as is plausible, the effective
exterior
volume $V$ is not much bigger that the volume of the bubble $v$, a
larger
contribution results.  Indeed, a careful discretized version of the
momentum
integrals in Eq.~(\ref{twophoton}) gives in general for the factor
multiplying the delta function in Eq.~(\ref{prob})
$v(1+v/V)^2$. The interference is {\em
constructive}, not destructive as I erroneously
claimed in my Leipzig talk, and negligible
as $V\to\infty$.  Taking the latter limit (but remembering that there
might
be up to a factor of 4 enhancement), and, appropriate for
$\tau_c/\tau_o
\gg1$, replacing $\sin^2\omega\tau_c\to1/2$, we obtain the
probability
of emitting a pair with momenta $\bf k$ and $\bf -k$ just as given by
Schwinger \cite{js3} (this now includes the equal contribution from
the
magnetic modes):
\begin{equation}
P_{\gamma\gamma}=v{(d{\bf
k})\over(2\pi)^3}\left(\epsilon-1\over4\right)^2,
\quad |\epsilon-1|\ll1.
\end{equation}
[For $|\epsilon-1|$ not small, Schwinger \cite{js3} generalized this
to
\begin{equation}
P_{\gamma\gamma}=2v{(d{\bf
k})\over(2\pi)^3}\ln{\epsilon^{1/4}+\epsilon^{-1/4}
\over2}.
\end{equation}
The numerical effect of this correction is not significant for a
first
estimate.]
The total number of photon pairs emitted is then, if dispersion is
ignored,
\begin{equation}
N=\left(4\pi\over3\right)^2\left(R\over\Lambda\right)^3
\left(\epsilon-1\over4
\right)^2,
\label{nophoton}
\end{equation}
where the cutoff wavelength is given by $K=2\pi/\Lambda$.  Such a
divergent
result should be regarded as suspect.\footnote{Although it is not
clear how
this is to be related to the divergent energy (\ref{casbulk2}),
Schwinger obtained both in Ref.~\cite{js3} as the imaginary and real
parts,
respectively, of a complex action.}
  It was Eberlein's laudable goal
\cite{eberlein} to put this type of argument on a sounder footing.
Nevertheless, if we put in plausible numbers, $\sqrt{\epsilon}=4/3$,
$R=4\times
10^{-3}$ cm, and, as in Schwinger's earlier estimate,
$\Lambda=3\times
10^{-5}$ cm, we obtain the required $N\sim 10^6$ photons per flash.

The problem with this estimate is one of time and length scales---for
the
instantaneous approximation to be valid, the flash time $\tau_f$ must
be
much less than the period of optical photons, $\tau_o\sim10^{-15}$ s.
This is consistent with the discussion in \S2, and acknowledged by
Liberati et al.\cite{visser}  On the other hand, the collapse time
$\tau_c\sim
10^{-5}$ s is vastly longer than $\tau_f$, and is therefore totally
irrelevant to the photon production mechanism.  The flash occurs near
minimum
radius, and thus the appropriate value of $R$ in Eq.~(\ref{nophoton})
would seem to be at least an order of magnitude smaller, $R\sim
10^{-4}$ em.
This would lead to $N<10^3$ photon pairs, totally insufficient.

\section{Conclusions}
We conclude by stating that the Casimir model fo sonoluminescence
remains
`unproven.'  The static Casimir effect can be applied only in the
adiabatic
approximation, where it seems clearly irrelevant.  The instantaneous
approximation grafted onto static configurations seems logically
deficient,
and again numerically irrelevant unless implausible parameters are
adopted.  What is still needed is a dynamical calculation of the
Casimir
effect.  The burden of proof is on the proponents of this mechanism.
\section*{Acknowledgments}
I thank Iver Brevik, Gabriel Barton, and Michael Bordag for useful
conversations.  This work was supported in part by a grant from the
U.S. Department of Energy.

\end{document}